\documentclass[cameraready]{Interspeech}

\title{Speaker Verification with Speech-Aware LLMs: Evaluation and Augmentation}

\author[affiliation={1,2}, orcid=0000-0001-8953-7872, correspondingauthor]{Thomas}{Thebaud}
\author[affiliation={1}, orcid=0009-0008-4451-1319]{Yuzhe}{Wang}
\author[affiliation={1}, orcid=0000-0002-3033-7005]{Laureano}{Moro-Velazquez}
\author[affiliation={1,2}, orcid=0000-0001-9459-8426]{Jesus}{Villalba-Lopez}
\author[affiliation={1,2}, orcid=0000-0002-4489-5753]{Najim}{Dehak}

\address{$^1$ Electrical and Computer Engineering Department, Johns Hopkins University, Baltimore, MD, USA\\
         $^2$ Human Language Technology Center of Excellence, Johns Hopkins University, Baltimore, MD, USA}

\email{tthebau1@jhu.edu}


\usepackage{comment}
\usepackage{xcolor}
\usepackage{booktabs}
\usepackage{multirow}

\begin{document}

\maketitle

\begin{abstract}
Speech-aware large language models (LLMs) can accept speech inputs, yet their training objectives largely emphasize linguistic content or specific fields such as emotions or the speaker's gender, leaving it unclear whether they encode speaker identity. 
First, we propose a model-agnostic scoring protocol that produces continuous verification scores for both API-only and open-weight models, using confidence scores or log-likelihood ratios from the Yes/No token probabilities. 
Using this protocol, we benchmark recent speech-aware LLMs and observe weak speaker discrimination (EERs above 20\% on VoxCeleb1). 
Second, we introduce a lightweight augmentation that equips an LLM with ASV capability by injecting frozen ECAPA-TDNN speaker embeddings through a learned projection and training only LoRA adapters. 
On TinyLLaMA-1.1B, the resulting ECAPA-LLM achieves 1.03\% EER on VoxCeleb1-E, approaching a dedicated speaker verification system while preserving a natural-language interface.
\end{abstract}

\section{Introduction}

Large Language Models (LLMs)~\cite{liu2026ministral, touvron2023llama, openai2024gpt4technicalreport} have recently become essential components of modern artificial intelligence systems. 
They have evolved into versatile architectures capable of processing inputs and outputs from multiple modalities, such as images~\cite{liu2026ministral, shu2023audio} and speech~\cite{tang2023salmonn, chu2023qwen, li2024audio}. 
Speech-aware LLMs, defined by their capacity to process audio inputs~\cite{arora2025landscape}, extend traditional text-based transformers by incorporating acoustic front-ends or audio tokenizers, enabling them to process spoken language directly rather than relying solely on automatic speech recognition transcripts. 
This innovation opens the door to models that could reason jointly over linguistic content and paralinguistic cues.
However, most speech-aware LLMs either focus on linguistic understanding~\cite{shon2022slue,wang2025audiobench}, or audio-based question answering~\cite{lin2026audiorag, wang2025audiobench, ma2025mmar}, with a few prosody-based closed-set classification tasks such as accent, gender and emotion recognition~\cite{wang2025audiobench}.

Automatic Speaker Verification (ASV) aims to determine whether two speech recordings originate from the same speaker. 
It is a critical component in numerous applications, including biometric authentication, but also personalized assistants and dialogue analysis. 
State-of-the-art ASV systems~\cite{desplanques2020ecapa, tu2022survey} are usually based on speaker embeddings ($x$-vectors~\cite{snyder2018x}), which encode the identity of a speaker within a single vector.
These systems are highly optimized for identity discrimination and achieve remarkable performance on benchmarks such as VoxCeleb~\cite{hechmi2021voxceleb}. 
If they can be leveraged for a variety of purposes, from speech emotion recognition~\cite{pappagari2020x} to healthcare~\cite{favaro2023multilingual,hom2021application} or speaker characterization~\cite{kwasny2021explaining}, they are generally narrow in scope: they do not reason over linguistic content, nor are they designed to integrate seamlessly with higher-level reasoning tasks.

The emergence of speech-aware LLMs raises an intriguing question: can a single large-scale, general-purpose model also perform fine-grained biometric tasks such as speaker verification? More specifically, do these models internally encode sufficient speaker-discriminative information, and can this information be harnessed or enhanced through appropriate training strategies? If so, this would suggest a path toward unified architectures capable of both high-level reasoning and low-level acoustic discrimination, reducing the need for task-specific pipelines.

In this work, we investigate the capabilities of current speech-aware LLMs for automatic speaker verification. We analyze whether their learned representations contain discriminative information for speaker identity. 
Building on this analysis, we propose a fine-tuning strategy that augments two widely used open-weight small-scale LLMs, tinyLLaMA 1.1B~\cite{zhang2024tinyllama} and Ministral3 3.3B~\cite{liu2026ministral}, with speaker verification capabilities by utilizing a pre-trained ECAPA-TDNN network from the speech brain toolkit~\cite{speechbrain}. 
Our approach adapts these models to produce speaker-discriminative representations while preserving their general modeling capacity, effectively creating '\textit{speaker-aware}' LLMs.

Our contributions are as follows:
\begin{itemize}
    \item We propose a model-agnostic protocol to evaluate speaker verification capability in speech-aware LLMs.
    \item We show that off-the-shelf speech-aware LLMs exhibit weak speaker discrimination on VoxCeleb1, mainly relying on coarse speaker characteristics.
    \item We introduce a lightweight augmentation that injects frozen ECAPA speaker embeddings with LoRA adaptation to equip LLMs with ASV capability, achieving near SOTA capabilities.
\end{itemize}

\section{Related Works}
\subsection{Automatic Speaker Verification}
Automatic Speaker Verification (ASV) has been extensively studied over the past two decades, evolving from generative statistical models to highly discriminative deep learning systems.
Most of those systems have been evaluated on the Voxceleb1 dataset~\cite{nagrani2020voxceleb}, a corpora of over a 100k utterances extracted from YouTube videos of 1251 celebrities, which contains 3 testing splits: Original, Extended and Hard (Vox1-O/E/H).

The introduction of the i-vector~\cite{dehak2010front} marked a major milestone, providing a compact representation of speaker characteristics that could be compared using probabilistic linear discriminant analysis (PLDA), which showed a 8.8\% EER on Vox1-O trained under the same conditions~\cite{nagrani2020voxceleb}.
The $x$-vector framework~\cite{snyder2018x} replaced generative factor analysis with a deep neural network trained to classify speakers, extracting fixed-dimensional embeddings from variable-length utterances, which bought up to 44\% relative improvement in EER in various scenarios evaluated on SITW~\cite{mclaren2016speakers}. 
Subsequent work introduced architectural refinements and training objectives tailored to speaker discrimination. In particular, time-delay neural networks (TDNNs) and their enhanced variants, such as ECAPA-TDNN~\cite{desplanques2020ecapa}, incorporate channel attention, multi-scale feature aggregation, and squeeze-and-excitation mechanisms to better capture long-range speaker characteristics, which pushed the EER down to 0.8\% on Vox1-O.
If more recent improvements have been produced since, we selected ECAPA-TDNN for our experiments for its accessibility through the speechbrain toolkit~\cite{speechbrain}, maximizing the reproducibility of our results.

\subsection{Speech-Aware Large Language Models}
Large Language Models (LLMs) are transformer-based architectures trained on large-scale text corpora using self-supervised next-token prediction.
This scaling has led to strong reasoning abilities, making LLMs a core component in many frameworks.
Open-weight models such as \textit{LLaMA}~\cite{touvron2023llama} and \textit{Ministral3}~\cite{liu2026ministral} provide efficient decoder-only architectures with competitive generalization, while proprietary systems such as \textit{GPT}~\cite{openai2024gpt4technicalreport} and \textit{Gemini}~\cite{team2023gemini} extend these capabilities to multimodal reasoning.
Although originally designed for text, transformers are modality-agnostic and can process other inputs once mapped into a compatible embedding space.

For speech, raw waveforms are typically encoded using pretrained acoustic models such as HuBERT~\cite{hsu2021hubert}, WavLM~\cite{chen2022wavlm}, or Whisper~\cite{radford2023robust}, or discretized using speech tokenizers and neural codecs~\cite{zhang2023speechtokenizer,dac, encodec}.
Speech-aware LLMs integrate these representations with pretrained language models through learned projection layers or adapters~\cite{thebaud2025enhancing, fang2024llama, tang2023salmonn, chu2023qwen, ding2025kimi}, enabling tasks such as speech recognition, spoken QA, and multimodal dialogue~\cite{arora2025landscape}.
However, their training objectives primarily target linguistic and semantic understanding, leaving open the question of whether they encode sufficiently discriminative speaker information for biometric tasks such as speaker verification.
We will investigate some of those models, utilizing a range of architectures, such as \textit{Qwen-2.5-7B}~\cite{chu2023qwen} and \textit{AudioFlamingo3}~\cite{goel2025audio} which leverages a Whisper encoder for its audio inputs or \textit{Kimi-audio-7B}~\cite{ding2025kimi} which also adds a neural audio codec for noise and music processing. 

\vspace{-3mm}
\section{Methods}
\vspace{-2mm}
\subsection{Dataset}
\label{sec:data}
VoxCeleb1~\cite{nagrani2020voxceleb} and VoxCeleb2~\cite{chung2018voxceleb2} are large-scale, publicly available audiovisual datasets widely used for speaker recognition research. 
VoxCeleb1 contains over 100,000 utterances from 1,251 celebrities, extracted from YouTube interview videos. 
VoxCeleb2 significantly expands this effort, comprising over one million utterances from 5,994 speakers. 
When training is needed, the experiment performed in this article will use the development set of VoxCeleb2 as training, and the testing set of VoxCeleb2 as validation. Both sets are defined in the original article and contain disjoint speakers.
For ablation studies, we also define a smaller set of VoxCeleb2-dev, using only 10\% of the speakers selected at random, and keeping only 10 utterances per speaker, which we name \textit{VoxCeleb2-dev-XS}, containing 600 speakers, for 6k utterances and a total of 12.4h of audio.
The testing sets used for all experiments will be the 3 test splits from VoxCeleb1 mentioned previously: the Original, Extended, and Hard trials, which are defined as lists of pairs of enroll and test utterances to compare.

    

\subsection{Proposed Speaker-Aware LLM architecture}
\begin{figure*}
    \centering
    \vspace{-5mm}
    \includegraphics[width=\linewidth]{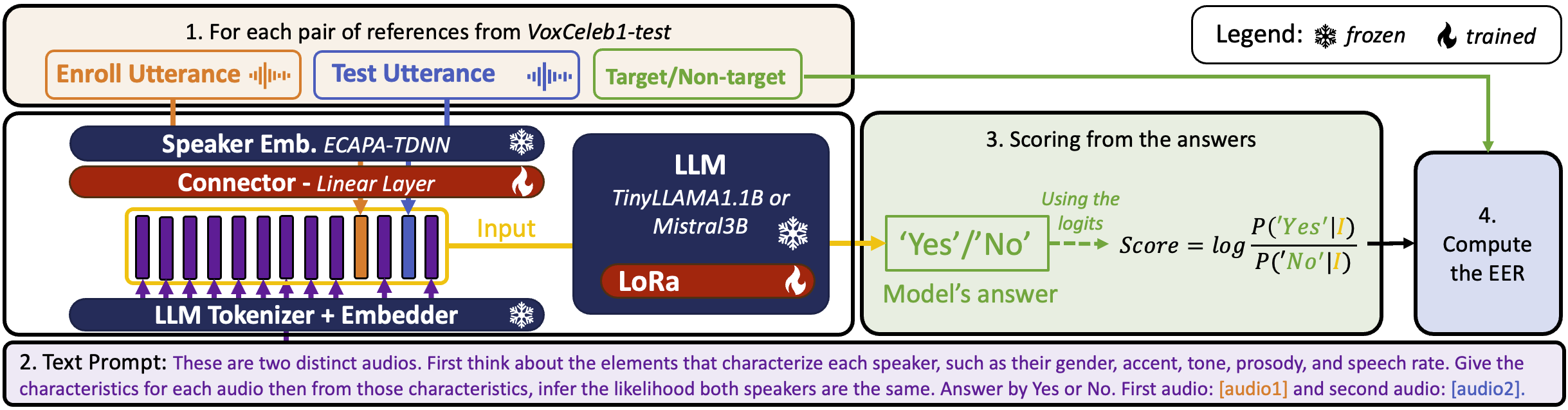}
    \caption{Schematic of the pipeline to train and test a speaker-aware LLM.}
    \label{fig:archi}
    \vspace{-5mm}
\end{figure*}

Cascaded \textit{speech}-aware LLMs are usually built around a pretrained speech encoder, a connector, and a pretrained LLM, which are finetuned to work jointly.
We propose a cascaded \textit{speaker}-aware LLM, as shown in Figure \ref{fig:archi}, which is built using a pretrained ASV system, a connector, and a pretrained LLM:
\begin{itemize}
    \item The ASV system is an ECAPA-TDNN~\cite{desplanques2020ecapa}, trained on VoxCeleb2-dev using speechbrain toolkit~\cite{speechbrain}, which shows 0.89\%, 0.45\% and 0.96\% EER respectively on VoxCeleb1 Original, Extended and Hard splits using cosine scoring. This system will be frozen during training.
    \item The connector is a linear layer, which is used to project the x-vectors in the dimension of the text embeddings of the LLM.
    \item The LLM is either a TinyLLaMA 1.1B~\cite{zhang2024tinyllama} or a Ministral3 3.3B~\cite{liu2026ministral}. Each model is finetuned using Lora adaptors~\cite{hu2021lora}.
\end{itemize}
The two Speaker-Aware configurations will be referred as SA-TinyLLaMA and SA-Ministral3 depending on the LLM used.

\subsection{Metrics}
ASV is a binary classification task, where a pair of utterances is either from the same speaker (target) or not (non-target).
Standard evaluation of an ASV system relies on the computation of a likelihood score for a predefined list of target and non-target pairs of utterances, which is used to compute the Equal Error Rate (EER).

This metric requires a continuous set of scores for evaluation.
If open-source models allow access to intermediate representations and the likelihood of an answer for different tokens, closed-source models usually only output the text answer to a query, which makes the computation of a continuous score a harder task. 

The next section details the proposed techniques to obtain a score from speech-aware LLMs

\subsection{Scoring speaker awareness in Speech-aware LLMs}
\subsubsection{Confidence scoring based on speaker characteristics}
\label{sec:score_conf}
In the case of closed-weight LLMs, or when the logits are not available, the only available output is the text.
We propose to ask for a confidence score between 0 and 100 in the probability that two given utterances are from the same speaker.
This confidence score can then be used to compute the EER.
The proposed prompt is: 

\noindent\fbox{%
\begin{minipage}{0.98\columnwidth}
{\scriptsize\ttfamily\raggedright\tiny
These are two distinct audios.
First, think about the elements that characterize each speaker, such as their gender, accent, tone, prosody, and speech rate.
Give the characteristics for each audio
Then, from those characteristics, infer the likelihood that both speakers are the same.
Answer by Yes or No, and give a confidence score between 0 and 100:  
0 corresponds to the certainty that they are from different speakers, 
100 corresponds to the certainty that they are from the same speaker, 
And 50 means you are uncertain. \par
First audio: $[audio 1]$ and second audio: $[audio 2]$.\par
}
\end{minipage}%
}

\subsubsection{Log-likelihood scoring based on logits}
\label{sec:score_log}

For more open-weights LLMs, were the logits associated to an answer are accessible, we can simplify the problem by asking a binary output, and compute the ratio of the logits for each answer.
The simplified prompt would be: 
\noindent\fbox{%
\begin{minipage}{0.98\columnwidth}
{\scriptsize\ttfamily\raggedright\tiny
These are two distinct audios.
First, think about the elements that characterize each speaker, such as their gender, accent, tone, prosody, and speech rate.
Give the characteristics for each audio
Then, from those characteristics, infer the likelihood that both speakers are the same. 
Answer by Yes or No. \par
First audio:$[audio 1]$ and second audio:$[audio 2]$.\par
}\end{minipage}}

Then, we can access the logits probability for the token \texttt{Yes} and the logits for the token \texttt{No}, and compute a log likelihood ratio as:
\begin{equation}
    LLR(prompt) = log(\frac{p(\texttt{Yes}|prompt)}{p(\texttt{No}|prompt)})
\end{equation}
This $LLR$ can be used subsequently as a score to compute the EER, offering a much finer-grained analysis than the integer confidence score previously used.
This technique is used to evaluate the capabilities of our two proposed cascaded speaker-aware LLMs: SA-TinyLLaMA and SA-Ministral3.

\subsection{Experiments}
\label{sec:experiments}

\subsubsection{Off-the-shelf speech-aware LLMs: confidence scoring}
\label{sec:exp_conf}

We probe a set of off-the-shelf speech-aware LLMs using the confidence-scoring protocol described in Section~\ref{sec:score_conf}.
For each VoxCeleb1 trial (enrollment utterance, test utterance) we issue a single prompt containing the two audio segments and request a binary decision (same/different speaker) together with a confidence score in $[0,100]$.
The returned confidence is used as the trial score to compute EER on the VoxCeleb1 trial lists (Vox1-O/E/H).

This protocol is used to evaluate GPT-4.0-audio~\cite{openai2024gpt4technicalreport} (model \texttt{gpt-4o-audio-preview-2025-06-03}), Qwen-2.5-7B~\cite{chu2023qwen}, Gemini 3-flash and 2.5-flash-lite\cite{team2023gemini}, AudioFlamingo3~\cite{goel2025audio}, and Kimi-audio-7B~\cite{ding2025kimi}.

\noindent\textbf{Failure rate:} In addition to verification performance, we track the fraction of trials for which the model response cannot be parsed into a valid decision and confidence score (reported as \emph{failure rate}).

\begin{table*}[ht]
    \centering
    \vspace{-3mm}
    \caption{Speaker-awareness evaluation of a set of Speech-aware LLMs using confidence scoring. EER (\%) is shown for each split of Voxceleb1-test, as well as the confidence score failure rate, accuracy and failure rate for gender and accent prediction. Accuracies are left empty for Kimi as those fields were never predicted. Note that in case of failure to provide a score for a trial, the trial is ignored in the computation of the EER.}
    \vspace{-3mm}
    \resizebox{\linewidth}{!}{
    \begin{tabular}{l c c c c c c c c}
    \toprule
    Model & \multicolumn{3}{c}{EER $\%\downarrow$}  & Failure & Gender  &  Gender & Accent  &  Accent\\
            &  Vox1-O & Vox1-E & Vox1-H                 & Rate $\%\downarrow$    & Accuracy $\%\uparrow$ & Predicted $\%\uparrow$ & Accuracy $\%\uparrow$ & Predicted $\%\uparrow$\\
    \midrule
     Qwen-2.5-7B~\cite{chu2023qwen}                         & 37.01 & 34.83 & 45.43    & 0.68  & \textbf{97.98} & 99.82 & 75.45 & 76.76 \\
     Kimi-audio-7B~\cite{ding2025kimi}                      & 43.58 & 43.02 & 43.93    & 16.12 & -       & 0     & -       & 0     \\
     Gemini3-flash~\cite{team2023gemini}                    & 45.13 & 45.64 & 44.26    & \textit{0.38}  & 92.16 & \textit{99.75} & \textbf{84.99} & \textbf{85.72} \\
     Gemini2.5-flash-lite~\cite{team2023gemini}             & 36.15 & 37.09 & 47.55    & 16.56 & 91.40 & \textbf{99.91} & \textit{83.32} & \textit{83.90} \\
     GPT4.0-audio~\cite{openai2024gpt4technicalreport}      & \textbf{22.62} & \textbf{21.88} & \textit{38.91}  & \textbf{0.05} & \textit{97.32} & 99.42 & 82.65 & 82.90 \\
     AudioFlamingo3~\cite{goel2025audio}                         & \textit{32.90} & \textit{31.00} & \textbf{31.51}    & 76.23 & 77.29 & 55.68  & 59.06 & 47.95 \\
    \bottomrule
    \end{tabular}
    }
    \vspace{-5mm}
    \label{tab:results_confidence}
\end{table*}

\subsubsection{Unprompted speaker characterization}
As the prompt requests the model to consider paralinguistic elements such as the accent, gender, and prosody of the speaker, most models explicitly detail their perception of those qualities.
We probe the outputs to measure the percentage of outputs containing gender and accents from the speakers, and report both the frequency of reporting and its accuracy. 

\noindent\textbf{Gender Metrics:} To simplify the evaluation process, only the outputs '\texttt{male}' or '\texttt{female}' are considered as reported gender\footnote{No other expression of gender was noted in the outputs of any model.}. 
The gender accuracy is computed using the gender labels provided in the metadata.
According to the metadata, speakers in VoxCeleb1 are 44.84\% female and 55.15\% male.

\noindent\textbf{Accent Metrics:} Accent accuracy is computed using the nationality labels provided in the metadata.
Attribution of a speaker to any geographical location is considered as a successful accent prediction.
If the model predicts a more restrictive label than the nationality ('\texttt{London accent}' or '\texttt{Scottish accent}' for '\textit{UK}' for example), the accent is counted right.
A less restrictive accent ('\texttt{Hispanic accent}' for '\textit{Mexico}' for example) is counted as wrong.
According to the metadata, speakers in VoxCeleb1 are 63.86\% from the USA, 17.18\% from the UK, 4.31\% from Canada, and other 33 nationalities are each less than 3\%.

\vspace{-3mm}
\subsubsection{Speaker-aware LLMs training and evaluation}
\label{sec:exp_llr}

\noindent\textbf{Training:}
All the proposed speaker-aware models are trained for 50 epochs on VoxCeleb2-dev, using a batch size of 64, a learning rate of $10^{-4}$ for all trainable parts, using VoxCeleb2-test as a validation set, on a single Nvidia A100 80Gb GPU.
The models are trained for next token prediction, to predict either '\texttt{Yes}' or '\texttt{No}', using batches composed of half target pairs and half non-target pairs.
The model with the best validation EER across epochs is kept.
The training and testing code for those models is available at this address\footnote{\url{https://github.com/thomasthebaud/ASV-with-SpeechLLMs}}.

\noindent\textbf{Evaluation:}
We evaluate our speaker-augmented models using the log-likelihood scoring protocol from Section~\ref{sec:score_log}, where the verification score is the log-likelihood ratio (LLR) between the \texttt{Yes} and \texttt{No} tokens, and present the results for the 3 evaluation split of VoxCeleb1.

\begin{table}[ht]
    \centering
    \caption{EER (\%) results on the splits of VoxCeleb1 for our proposed speaker-augmented LLMs, evaluated using log-likelihood based scoring. The ECAPA-TDNN system is shown as a comparison, evaluated using cosine scoring.}
    \vspace{-3mm}
    \resizebox{0.95\linewidth}{!}{
    \begin{tabular}{l l l l }
    \toprule
    Model & Vox1-O & Vox1-E & Vox1-H \\
    \midrule
     ECAPA-TDNN\cite{desplanques2020ecapa} & 0.89 & 0.45 & 0.96 \\
    \midrule
     SA-Ministral3          & 14.76 &  15.88  & 21.04 \\
     SA-TinyLLaMA           & \textbf{1.87}  & \textbf{1.03}    & \textbf{2.20} \\
     SA-TinyLLaMA$^F$       & 5.48  &  4.21   &  6.60  \\ 
     SA-TinyLLaMA$_{XS}$    & 3.57  &  2.21   &  3.44  \\
     SA-TinyLLaMA$_{XS}^F$  & 27.01 & 27.82   & 28.55  \\
    \bottomrule
    \end{tabular}
    }
    \vspace{-5mm}
    \label{tab:results_log}
\end{table}

\noindent\textbf{Ablations:}
We train a variant where the LLM backbone remains frozen (no LoRA adapters) and only the connector is learned.
We denote this model SA-TinyLLaMA$^F$.
This setting follows the frozen-LLM adaptation proposed in~\cite{thebaud2025enhancing}, which reported 12.08\% EER on VoxCeleb1-O using LLaMA2 3B.

Considering this problem could be seen merely as an alignment problem between the speaker embedding space and the LLM embedding space, we hypothesize that only a fraction of the speakers would be enough.
To that end, we train a model using the VoxCeleb2-dev-XS subset defined in section \ref{sec:data}, and name this model SA-TinyLLaMA$_{XS}$.
To complete the ablation, a model is trained on the VoxCeleb2-dev-XS subset with a frozen TinyLLaMA, named SA-TinyLLaMA$_{XS}^F$.

\vspace{-2mm}
\section{Results}
\vspace{-1mm}

\subsection{Off-the-shelf speech-aware LLMs}
\label{sec:results_off_the_shelf}
\vspace{-1mm}

Table~\ref{tab:results_confidence} shows that off-the-shelf speech-aware LLMs exhibit weak speaker discrimination under the confidence-scoring protocol.
EERs remain far above a dedicated ASV system, ranging from 22.62\% (GPT-4.0-audio on Vox1-O) to approximately 45\% (Gemini), with several models operating close to chance level (50\% EER).
These results indicate that, under standard instruction-following prompts, current speech-aware LLMs do not reliably expose speaker-discriminative information sufficient for verification, even when explicitly asked to reason about paralinguistic cues.
Those poor performances could be explained by the coarse granularity of the scores, effectively allowing for only 101 levels, but those are not even used by the models, as all the prompted models yielded up to 16 different scores\footnote{Full set of outputed scores across all models: [0, 1, 2, 5, 10, 15, 20, 30, 40, 45, 50, 55, 60, 65, 70, 75, 80, 85, 90, 95, 99, 100]}.
We also observe substantial differences in robustness across models.
In particular, AudioFlamingo3 exhibits a high failure rate (76.23\%), making its raw EER difficult to interpret.

\vspace{-3mm}
\subsubsection{Speaker Characteristics prediction}
\label{sec:results_paralinguistics}

Although verification performance is poor, several models are nevertheless able to infer coarse speaker attributes.
For example, Qwen-2.5-7B, Gemini, and GPT-4.0-audio achieve high gender classification accuracy when a gender label is produced (92-98\% in Table~\ref{tab:results_confidence}), with near-complete gender coverage.
This contrast reveals a clear limitation: although the models capture coarse speaker attributes, such information is insufficient for fine-grained speaker identity discrimination. 
This is particularly evident on Vox1-H, where the absence of cross-gender trials reduces the usefulness of gender cues and leads to a noticeable performance drop. 
These findings support the hypothesis that current speech-aware LLM training objectives prioritize linguistic and high-level paralinguistic features rather than identity-specific representations.

\vspace{-3mm}

\subsection{Speaker-aware LLMs results}
\label{sec:results_augmented}

In contrast to off-the-shelf speech-aware LLMs, Table~\ref{tab:results_log} shows that injecting frozen ECAPA speaker embeddings and training only a lightweight connector together with LoRA adapters yields a large improvement in verification performance.
SA-TinyLLaMA achieves EERs close to the ECAPA-TDNN cosine baseline, indicating that a general-purpose LLM can be endowed with strong ASV capability when provided with an explicit speaker representation and minimal task-specific adaptation, while preserving a natural-language interface.

Ablation results further highlight the importance of adapting the LLM backbone.
When the LLM is frozen, and only the connector is learned, performance degrades substantially: SA-TinyLLaMA$^{F}$ reaches 5.48\% EER on Vox1-O.
This confirms that the gains are not solely attributable to the quality of the injected speaker embeddings; parameter-efficient adaptation is necessary for the LLM to reliably interpret the speaker representation and produce stable verification decisions under the prompted output format.
Finally, we observe that the smaller TinyLLaMA-1.1B backbone outperforms larger backbones (LLaMA-3B and Ministral3-3B) in our current training setup, an effect that merits further investigation.
The poor performance of Ministral3-3B remains unexplained despite efforts to optimize its training, but it may be attributed to differences in model size and fine-tuning procedures, which could require alternative loss functions for optimal performance.
\vspace{-2mm}
\section{Conclusion}
This work investigated whether modern speech-aware LLMs encode speaker identity information and whether this can be leveraged for automatic speaker verification (ASV).
We introduced a model-agnostic evaluation protocol that derives continuous verification scores from both API-only systems (via prompted confidence) and open-weight models (via log-likelihood ratios over \texttt{Yes}/\texttt{No} tokens).
Benchmarking on VoxCeleb1 shows that off-the-shelf speech-aware LLMs exhibit weak speaker discrimination, with EERs typically above 20\%.

We proposed a lightweight augmentation that injects frozen ECAPA-TDNN speaker embeddings into an LLM through a learned projection and LoRA adaptation.
On TinyLLaMA-1.1B, the resulting model approaches the performance of a dedicated ECAPA-TDNN system while preserving a natural-language interface.

Overall, our results suggest a practical path toward speaker-aware LLMs by explicitly integrating strong speaker representations rather than relying on implicit learning.
Nevertheless, our evaluation is limited by the coarse and model-dependent nature of confidence-based scoring for closed systems, as well as by parsing failures in some APIs, which constrain direct comparisons. 
If this article show the technical possibility to augment LLMs with speaker verification capabilities, the computing cost both at training and inference remains orders of magnitude higher than traditional dedicated systems.
Future work will explore more robust scoring strategies and extend this framework toward temporally resolved speaker modeling, enabling tasks such as diarization and multi-talker conversation analysis within speech-aware LLMs.

\newpage

\section{Generative AI Use Disclosure}
Parts of this manuscript were edited and rewritten for clarity, grammar, and style with the assistance of a generative AI language model. 
The model was used only for language refinement and formatting suggestions; it did not contribute to the scientific content of this work. 
In particular, no ideas, hypotheses, experimental designs, implementations, code, datasets, analyses, results, or bibliographic references were generated by any GenAI system. 
All technical decisions and the final text were reviewed and verified by the authors, who take full responsibility for the content.

\bibliographystyle{IEEEtran}
\bibliography{main}

\end{document}